\pdfoutput=1
\documentclass[a4paper,fleqn,usenatbib]{mnras}

\usepackage{newtxtext,newtxmath}

\usepackage[T1]{fontenc}
\usepackage{ae,aecompl}

\usepackage{graphicx}	
\usepackage{amsmath}	

\usepackage{color}
\usepackage{ulem}
\usepackage{natbib}

\title[Cloaking of a naked singularity]{Astrophysical cloaking of a naked singularity
}

\author[R. S. S. Vieira and W. Klu\'zniak]{Ronaldo S. S. Vieira\,$^{1}$\thanks{Corresponding authors}\thanks{ronaldo.vieira@ufabc.edu.br}
and W\l odek Klu\'zniak\,$^{2,3}$
\thanks{wlodek@camk.edu.pl}
\\
$^{1}$ Centro de Ci\^encias Naturais e Humanas, Universidade Federal do ABC, 09210-580 Santo Andr\'e, SP, Brazil\\
$^{2}$ Copernicus Astronomical Center, ul. Bartycka
  18, PL-00-716, Warszawa, Poland\\
$^{3}$ Institute of Physics, Silesian University in Opava, Bezru\v{c}ovo n\'am.~13, CZ-746\,01 Opava, Czech Republic.
}

\date{Accepted XXX. Received YYY; in original form ZZZ}

\pubyear{2023}

\begin{document}
\label{firstpage}
\pagerange{\pageref{firstpage}--\pageref{lastpage}}
\maketitle

\begin{abstract}
A massive naked singularity would be cloaked by accreted matter, and thus may appear to a distant observer as an opaque \mbox{(quasi-)}spherical surface of a fluid, not unlike that of a star or planet.
We present here analytical solutions for levitating atmospheres around a wide class of spherically symmetric naked singularities. Such an atmosphere can be constructed in every spacetime which possesses a zero-gravity radius and which is a solution of a (modified-)gravity theory possessing the usual conservation laws for matter. Its density peaks at the zero-gravity radius and the atmospheric fluid is supported against infall onto the singularity by gravity alone. In an astrophysical context, an opaque atmosphere 
would be formed in a very short time by accretion of ambient matter onto the singularity---in a millisecond for an X-ray binary, in a thousand seconds for a singularity traversing interstellar space, and a thousand years for a singularity that is the central engine of an AGN. 
\end{abstract}




\section{Introduction}

Ordinary stars, and many planets, are surrounded by a gaseous, sometimes liquid, envelope. While a density inversion may occur under certain circumstances, hydrostatic equilibrium in Newtonian gravity requires the pressure of these atmospheres to increase with depth down to their base, where they are supported by a denser medium. This is not always true in Einstein's gravity, as well as other (``modified'') theories of gravity.

Stellar atmospheres around spherical, luminous relativistic stars may levitate. In fact, it was shown \citep{abramowiczEtal10990ApJ} that, at a certain radius, matter at rest on an imaginary spherical surface around a near-Eddington luminosity neutron star
will be in stable equilibrium with respect to radial perturbations. Recently, the surface was termed the Eddington capture sphere (ECS), because strong radiative drag brings orbiting particles to rest on this surface, where the matter is supported against gravity by radiative forces \citep{ohEtal2010PhRvD, stahlEtal2012AA, stahlEtal2013AA}. {The restriction to strictly spherical geometry of the star may be crucial for the existence of ECS \citep{wielgus2019MNRAS}.}
If a quantity of dissipative fluid is placed in the vicinity of the ECS, the fluid shell will attain hydrostatic equilibrium. 
Since the density and pressure of such a shell drop off with distance to the ECS, whether outside it or inside it, one has an atmosphere well separated from the surface of the radiating star \citep{wielgusEtal2015MNRAS}. When geometrically thin, the atmosphere is nearly symmetric with respect to reflection in the ECS. Vibrational modes have been computed for such levitating atmosphere  \citep{bollimpalliEtal2019MNRAS}, and their frequencies depend on the properties of the central body (mass, radius, luminosity). 

What happens if we have a central naked singularity instead of a star? Naked singularities are hypothetical singularities in spacetime which are not ``covered'' by an event horizon, so in principle they are accessible to observations. In this sense, it would be possible to ``see'' the singularity. 
There are many examples of naked-singularity spacetimes in the literature, most of them coming from modified theories of gravity (e.g. \citealp{kovacsHarko2010PRD, vieiraMarekEtal2014PRD, katkaVieiraEtal2015GRG, boshkakayev2016PRD}). However, naked-singularity solutions also appear in general relativity (GR), such as the Chazy-Curzon solution, the $Q>m\,$ Reissner-Nordstr\"om solution, and the superspinning $a>1$ Kerr naked singularity solution \citep{griffithsPodolsky2009exact}.

It is a general property of spherically symmetric naked singularities that their gravitational field creates a region around them such that if a particle is dropped in this region, it will not fall into the singularity but instead will be repelled to larger radii \citep{vieiraMarekEtal2014PRD}. This region forms a spherical region of ``repulsive gravity'' or an ``antigravity region.''  On the other hand, far from the singularity gravity is attractive. The spherical interface between the two regions may be termed a zero-gravity sphere as it has the property that a test particle placed
there will follow a geodesic with constant spatial coordinates.  
 {Since for} naked-singularity spacetimes with the above property {there must be a minimum of the effective potential between the inner repulsive and outer attractive region,} this zero-gravity radius corresponds to a stable equilibrium point of the system. Moreover, this equilibrium is maintained by gravity alone, without the need of a radiative force (which is required in the case of the ECS, for instance). In this paper, we construct levitating atmospheres in the context of the antigravity region around naked singularities, and investigate their properties. 
To safely discuss atmospheric properties within the hydrodynamic approximation, we only consider fairly massive singularities, exceeding the mass of a large planetoid. In particular we consider the hypothesis that some X-ray sources may be powered by accretion onto a naked singularity of stellar mass in the case of X-ray binaries, and/or supermassive ones for AGNs.

We consider astrophysical naked singularities which must attract and accrete some ambient matter. Energy is conserved in purely geodesic motion, consequently in such motion test particles falling in from a larger radius would oscillate around the zero-gravity radius, but would never stop there. In reality, dissipative processes in the infalling matter will lead to it settling in the ``potential energy well'' of the gravitating object. For example, it is well known that in accretion disks  fluid particles in nearly circular orbits gradually diminish their radial distance from the gravitating object while the disk transports angular momentum away and radiates their excess energy through a dissipation mechanism (e.g., \citealp{pringle1981ARAA}). We expect that this or a similar mechanism will operate for matter present around the singularity, as it does for less exotic objects, such as neutron stars or black holes. Thus, some quantity of ambient matter will eventually settle down at the zero-gravity radius, forming a sphere of matter similar to the ECS\footnote{In the case of the ECS, even in the absence of dissipative fluid interactions, nearby orbiting particles settle inexorably  on that surface due to the action of the Poynting-Robertson effect \citep{stahlEtal2012AA, wielgusEtal2012AA}.}. Once matter collects on the zero-gravity sphere, it may generate an internal pressure, forming then a ``levitating atmosphere'' around the naked singularity, i.e., a stationary shell of fluid positioned some distance away from the singularity, on both sides of a spherical surface at the zero-gravity radius.

\begin{center}
	\begin{figure}
		\includegraphics[width = 0.98\columnwidth]{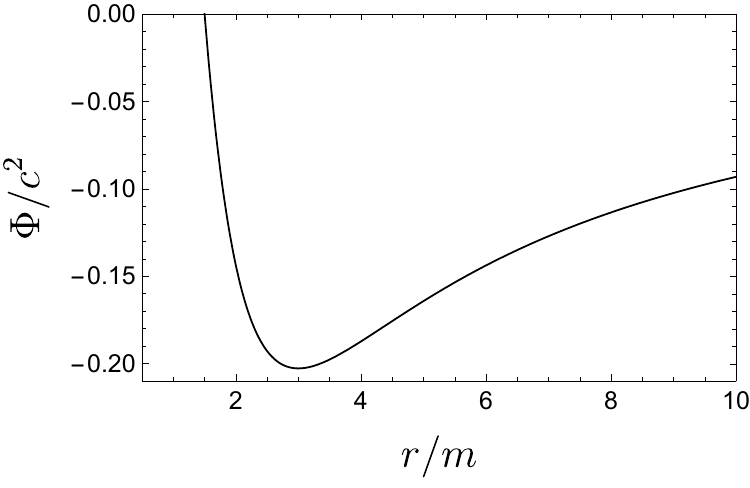}
		\caption{The dimensionless effective potential $\Phi(r)/c^2$ for Reissner-Nordstr\"om spacetime with $r_0/m=(Q/m)^2 = 3$}
		\label{fig:PhiRN}
	\end{figure}
\end{center}
\section{Singularities and cosmic censorship}
{There is no general solution to the problem of gravitational collapse of astrophysical bodies. The Cosmic Censorship conjecture \citep{penrose1969}, stating that the creation of a singularity will be accompanied by the formation of an event horizon hiding it from the external world, has never been proven as a theorem, and in fact the content and hypotheses of such a theorem were found to be difficult to formulate \citep[see, e.g., book by][]{joshiBook1993}. In fact, several classes of counter examples  have been found in GR, so the Cosmic Censorship conjecture cannot be true in general.}

{A brief history of key research results in collapse to a naked singularity can be found, e.g., in the introduction of the paper by \cite{giamboGiannoniMagliPiccione2004}. For a brief explanation of the mechanism of such collapse see \cite{joshiDadhichMaartens2002}.}

{If we confine ourselves to the astrophysically relevant collapse of a continuous matter distribution, the first examples of gravitational collapse to a naked singularity were found for dust, numerically \citep{eardleySmarr1979} and analytically \citep{christodoulou1984}. The 
conditions for dust collapse to naked singularities and those for collapse to a black hole are now known \citep{singhJoshi1996}.}

{For matter with non-zero pressure a breakthrough occurred with the pioneering work of \cite{oriPiran1990} who have shown how a self-similar fluid (i.e., one with pressure proportional to energy density) may collapse to a singularity. This work has been extended analytically to more general barotropic fluids by \cite{giamboGiannoniMagliPiccione2004}, and the authors identified the conditions under which naked singularities in the spherically symmetric collapse of non self-similar barotropic fluid are formed. For a more recent construction of collapse to a naked singularity in GR see \cite{joshiMalafarinaNarayan2011}.}

{It is clear that naked singularities can, in principle, be formed. It is not yet known how special (or conversely, general) the conditions must (may) be for the formation of a naked singularity in gravitational collapse of actually occurring matter distributions.}

{In this paper we will assume that naked singularities may have been formed, and we will explore some astrophysical consequences of their possible existence.}

\section{The zero-gravity sphere}
\label{sec:antigSphere}

Consider a general static, spherically symmetric metric
\begin{equation}
\label{eq:metric}
ds^2=-e^{2\Phi/c^2}\,c^2\,dt^2 + e^{2\Lambda/c^2}\,dr^2 + r^2\,(d\theta^2 + \sin^2\theta\,d\varphi^2)\,,
\end{equation}
where $\Phi$ and $\Lambda$ are functions of the radius:  $\Phi(r)$ and $\Lambda(r)$. In the Newtonian limit, $\Phi$ is the gravitational potential of the system. 

The 4-acceleration of an observer is given by $a_\mu = u^\alpha \nabla_\alpha u_\mu$, where $u^\mu$ is the observer's four-velocity. It can be shown that, for a static observer, $a_\mu = \partial_\mu\Phi$ \citep{semerakZellerinZacek1999MNRAS} for metric~(\ref{eq:metric}). Therefore the 4-acceleration of a static observer has only a radial component, $a_r = \Phi'(r)$. Then, if there exists a radius $r_0$ at which $\Phi'(r_0) = 0$, any static observer at this radius will be a geodesic observer and a test particle placed at this radius will remain at rest. This radius has often been called the `antigravity radius' \citep{vieiraMarekEtal2014PRD, katkaVieiraEtal2015GRG}. We will call it the `zero-gravity radius' and the corresponding 2-sphere the `zero-gravity sphere' of the spacetime. 

The zero-gravity sphere represents an equilibrium point for radial test-particle motion. This equilibrium will be stable if $\Phi''(r_0) > 0$ and unstable if $\Phi''(r_0) < 0$. Since we are interested in stable configurations, we will only consider the case $\Phi''(r_0) > 0$. In this case which, as remarked already, is quite common, a sufficient amount of matter can accumulate at $r_0$ that a balance between internal pressure and gravity will build up a shell of definite thickness.

A similar shell may be formed in the Schwarzschild metric only in the presence of radiation, owing to  relativistic redshift effects in luminosity---an effect which is absent in Newtonian gravity---that affect the balance between gravity and the radiative force on an ion, resulting in an equilibrium surface being present outside near-Eddington luminosity neutron stars   \citep{abramowiczEtal10990ApJ}. 
However, in the case presented here, the presence of the equilibrium surface is due to gravity alone, which produces a minimum of the effective potential for radial motion at the zero-gravity sphere.

\section{Levitating atmospheres around naked singularities}
\label{sec:atmospheres}

Consider a test perfect fluid whose energy-momentum tensor is given by
\begin{equation}
\label{eq:Tmunu}
T^{\mu\nu} =\frac{ (\varepsilon + p)}{c^2}\, u^\mu u^\nu + p\,g^{\mu\nu},
\end{equation}
with the fluid four-velocity $u^\mu$, energy density $\varepsilon$, and pressure $p$. Let us also  {consider a static observer with the normalized 4-velocity} $u^\alpha = c\,e^{-\Phi/c^2}\delta^\alpha_t$. Then the conservation laws $T^{\mu\nu}_{\ \ \ ;\nu} = 0$ (where a semi-colon denotes covariant differentiation) give us the general relativistic equation of hydrostatic equilibrium \citep{schutz2009book}
\begin{equation}
\label{eq:conservationLaw}
\frac{dp}{dr} = - \frac{\varepsilon + p}{c^2}\,\frac{d\Phi}{dr}.
\end{equation}
A remark may be made about the domain of validity of this equation and its applicability to modified theories of gravity.
The above equation 
has a different nature from the Tolman-Oppenheimer-Volkov (TOV) equation; it comes directly from the conservation laws. Therefore the static solutions of equation~(\ref{eq:conservationLaw}) give us the equilibrium configurations of test fluids in any theory of gravity in which matter satisfies the usual conservation laws $T^{\mu\nu}_{\ \ \ ;\nu} = 0$.
For our purposes, in which we consider only test fluids in a background metric of the form (\ref{eq:metric}), equation~(\ref{eq:conservationLaw}) completely determines the density and pressure profiles once we know the atmosphere's equation of state (EOS).
On the other hand, the TOV equation also takes into account Einstein's equations for the spherically symmetric case in general relativity. In this way, its generalization depends on the modified theory of gravity under consideration via its modified field equations, and would be important if we wanted to obtain self-gravitating atmospheres around the singularities. Our focus, however, is on the test-fluid approximation in which the atmosphere does not affect the background spacetime metric. 

If the background spacetime contains a zero-gravity sphere at $r=r_0$, with
a positive second derivative of the metric function, $\Phi''(r_0) > 0$, i.e., a minimum of the effective potential for radial motion (as in Fig.~\ref{fig:PhiRN}), then dissipation processes (whether in an accretion disk or Bondi accretion shock) will lead to a deposition of matter around this sphere, eventually forming a levitating atmosphere. 
In the case of a fluid, pressure gradients will rapidly spread the matter over the sphere. However, even dust settling down on this sphere would eventually fill its whole extension, because of the random nonradial motion, so that the central singularity will be cloaked by matter from all viewing angles. We discuss the opacity of the atmosphere in Section~\ref{sec:opacity}.

In the following, we suppose that the atmosphere is described by an ideal gas EOS,
$p = {k_\mathrm{B} T}\rho/{(\mu\, m_\mathrm{p})},$
where $T$ is the gas temperature, $k_\mathrm{B}$ is Boltzmann's constant, $m_\mathrm{p}$ is the proton mass, and $\mu$ 
is the mean molecular weight ($\mu = 0.5$ for ionised hydrogen).
We also assume that the atmospheric temperatures are relatively low, in the X-ray or low energy $\gamma$-ray range at most, so we can make the approximation $\varepsilon = \rho c^2$, $\rho$ being the atmosphere's rest-mass density. We also neglect the pressure term on the right-hand side of (\ref{eq:conservationLaw}) since it is then negligible
in comparison to the atmospheric energy density at low temperatures.  
In this case, equation~(\ref{eq:conservationLaw}) reduces to
\begin{equation}\label{eq:hydrostaicEquilibrium}
\frac{1}{\rho}\frac{dp}{dr} = - \frac{d\Phi}{dr},
\end{equation}
which is formally identical to the Newtonian equation of hydrostatic equilibrium, but it is important to keep in mind that $\Phi$ is a metric function in equation (\ref{eq:metric}) and not the Newtonian gravitational potential.


\begin{center}
	\begin{figure}
		\includegraphics[width = 0.98\columnwidth]{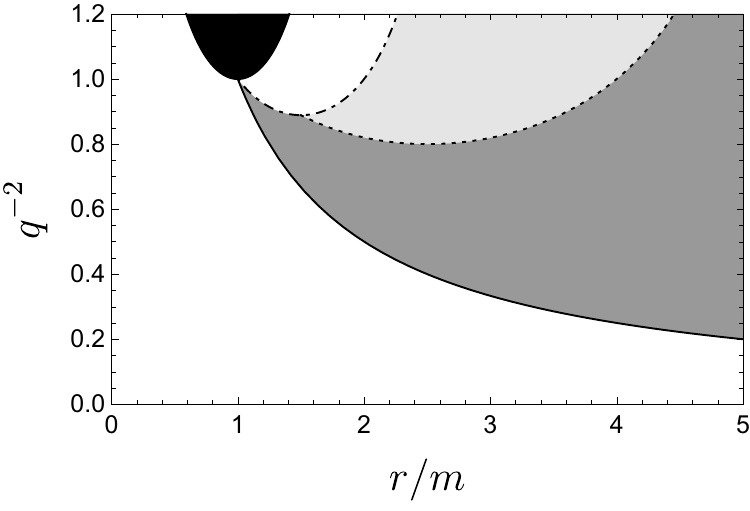}
		\caption{Stability diagram for Reissner-Nordstr\"om spacetime with  $q \equiv Q/m$. The solid black curve represents the zero-gravity radius. {\sl Black:} Region between horizons. {\sl Dark gray:}  Stability region for  {circular timelike} geodesics. {\sl Light gray:} Instability region for  {circular timelike} geodesics. {\sl White:} Region of no circular timelike geodesics. The dashed line denotes the locus of marginally stable orbit, the dot-dashed line the circular photon orbits.
		}
		\label{fig:stabRN}
	\end{figure}
\end{center}
\subsection{Geometrically thin approximation}

We now analyse the case in which the atmosphere is geometrically thin. Then we may expand the function $\Phi(r)$ around $r_0$ in equation~(\ref{eq:hydrostaicEquilibrium})  {through} second order in $(r-r_0)$ to obtain 
{$d\Phi/dr=(r-r_0)\Phi''(r_0)$ and, with the substitution of variables $d[(r-r_0)^2]=2(r-r_0)dr$},
\begin{equation}\label{eq:p--r-solve}
\frac{dp}{d[(r-r_0)^2]}= -\frac{1}{2} \Phi''(r_0)\, \rho\,.
\end{equation}
We see that, since we assume that geodesics in the zero-gravity sphere are stable against radial perturbations,  $\Phi''(r_0)>0$ (see e.g., Fig.~\ref{fig:PhiRN}), the pressure profile of the atmosphere has a peak at $r_0$ and decays with increasing distance from the sphere (on both sides), forming then an atmosphere which levitates at a finite distance from the singularity.

It must be noted that the above approximation is valid for rarefied atmospheres centered at the zero-gravity radius of any naked-singularity spacetime; the properties, given below in this Subsection, of the isothermal and polytropic atmospheres are valid regardless of the nature of the central naked singularity. In this regime of geometrical thinness of the atmosphere, it is not easily possible
to distinguish between different theories of gravity under consideration.
The specific spacetime enters in the atmospheric profile only via the zero-gravity radius $r_0$ and the constant $\Phi''(r_0)$.


\subsubsection{Isothermal atmosphere}

For an isothermal atmosphere of temperature $T$, we find that equation~(\ref{eq:p--r-solve})  
gives us a Gaussian profile for its pressure, peaked at $r=r_0$, 
\begin{equation}
p(r) = p_0 \exp\left[ -\frac{\mu\, m_\mathrm{p}}{\,2k_\mathrm{B} T\,}\,\Phi''(r_0)\,(r-r_0)^2 \right],
\end{equation}
whose half-width depends on the parameter $\Phi''(r_0)$. Here, $p_0 = p(r_0)$ is the pressure maximum. From the ideal gas equation, $\rho(r)$ will have the same functional form as $p(r)$.

\subsubsection{Polytropic atmosphere}

For a polytropic atmosphere $p = K\rho^\gamma$, with $\gamma>1$, equation~(\ref{eq:p--r-solve}) gives us the pressure profile, also peaked at $r=r_0$,
\begin{equation}
    p(r) = p_0\,\left\{ 1 - 
\frac{1-1/\gamma}{\,2K^{1/\gamma}\,(p_0)^{1-1/\gamma}\,}\, \Phi''(r_0)\,(r-r_0)^2 \right\}^{\gamma/(\gamma-1)},
\label{pthin}
\end{equation}
with $p_0 = p(r_0)$, and a corresponding density profile
\begin{equation}
\rho(r) = \rho_0\left\{ 1 - 
\frac{1-1/\gamma}{\,2K\,\rho_0^{(\gamma-1)}\,}\, \Phi''(r_0)\,(r-r_0)^2 \right\}^{1/(\gamma-1)},
\label{rhothin}
\end{equation}
where the density at the peak is $\rho_0 = \rho(r_0) = (p_0/K)^{1/\gamma}$.
The temperature profile is given by
\begin{equation}
	T = T_0 \left\{1 - \left(\frac{1-1/\gamma}{k_\mathrm{B} T_0/(\mu\, m_\mathrm{p})}\right)\cdot\frac{1}{2}\,\Phi''(r_0)\,(r-r_0)^2\right\}
\end{equation}
where 
\begin{equation}
    p_0/\rho_0 = k_\mathrm{B} T_0/(\mu\, m_\mathrm{p}).
    \label{cenT}
\end{equation}
The temperature maximum (central temperature) can be found by eliminating $p_0$ in the polytropic and ideal gas equations of state:
\begin{equation}
	 {k_\mathrm{B}}T_0= {\mu\, m_\mathrm{p}} K\rho_0^{(\gamma-1)}.
  \label{T0}
\end{equation}
Conversely, the polytropic constant can be expressed as
\begin{equation}
	  K= \rho_0^{(1-\gamma)}{k_\mathrm{B}}T_0/({\mu\, m_\mathrm{p}}).
	  \label{polconst}
\end{equation}


We note that, in general, once we assume a barotropic EOS $p(\rho)$ as a non-decreasing function of $\rho$, the pressure and density profiles will be both peaked at $r=r_0$, depending on radius via the expression $\Phi''(r_0)\,(r-r_0)^2$ regardless of the considered spacetime. 
Therefore, as anticipated above, in the geometrically thin approximation the atmospheres associated with a given barotropic EOS will have the same profile for every naked-singularity spacetime, the only difference between them being their width, associated with (inversely proportional to the square root of) $\Phi''(r_0)$.

\subsection{Atmospheric thickness}
\label{sec:thickness}

To assess the self-consistency of the  geometrically thin approximation we examine the thickness
of a polytropic atmosphere.

Rewriting equations~(\ref{pthin}), (\ref{rhothin}) in a convenient form, using equation~(\ref{polconst}),
\begin{equation}
p(r) = p_0\left\{ 1 - 
\frac{\gamma-1}{2\gamma}\, \frac{\mu\, m_\mathrm{p}\,c^2}{k_\mathrm{B}T_0} \frac{\Phi''(r_0)}{c^2}\,(r-r_0)^2 \right\}^{\gamma/(\gamma-1)},
\label{pthing}
\end{equation}
\begin{equation}
\rho(r) = \rho_0\left\{ 1 - 
\frac{\gamma-1}{2\gamma}\, \frac{\mu\, m_\mathrm{p}\,c^2}{k_\mathrm{B}T_0} \frac{\Phi''(r_0)}{c^2}\,(r-r_0)^2 \right\}^{1/(\gamma-1)},
\label{rhothing}
\end{equation}
we obtain an expression for the radii of the edges of the atmosphere $r_\pm$, where $\rho(r_\pm)=0$, and $r_-<r_0<r_+$,
\begin{equation}
\frac{|r_\pm -r_0|}{r_0}=
\left\{\frac{2\gamma}{\gamma-1}\, \left(\frac{k_\mathrm{B}T_0}{\mu\, m_\mathrm{p}\,c^2}\right) \left(\frac{c^2}{r_0^2\,\Phi''(r_0)}\right)\right\}^{1/2}.
\label{rminmax}
\end{equation}
Typically, ${r_0^2\,\Phi''(r_0)}/{c^2}\sim 1$ (equations~(\ref{eq:phibis}), (\ref{ibis})), so ${|r_\pm}-{r_0}|\ll r_0$ already for MeV temperatures. In the X-ray range ($k_\mathrm{B}T_0\sim 1\,$keV) the coefficient  $({k_\mathrm{B}T_0}/{\mu\, m_\mathrm{p}\,c^2})\sim 10^{-6}\ll 1$. 

Close to $r_0$ ($|r-r_0|\ll r_+-r_0)$ we can expand equation~(\ref{rhothin})
around the ``base'' of the atmosphere (at $r_0$)
to lowest order in $(r-r_0)^2$ as
\begin{equation}
\rho(r) \approx \rho_0\left\{ 1 - \frac{1}{2\gamma}\, \left(\frac{\mu\, m_\mathrm{p}\,c^2}{k_\mathrm{B}T_0}\right) \frac{\Phi''(r_0)}{c^2}\,(r-r_0)^2 \right\},
\label{rho}
\end{equation}
and equation~(\ref{pthin}) as
\begin{equation}
p(r) \approx p_0\left\{ 1 - \frac{1}{2}\, \left(\frac{\mu\, m_\mathrm{p}\,c^2}{k_\mathrm{B}T_0}\right) \frac{\Phi''(r_0)}{c^2}\,(r-r_0)^2 \right\}.
\label{pe}
\end{equation}
This parabolic profile of pressure and density close to $r_0$ is reminiscent of the corresponding profiles close to the midplane of an accretion disk\footnote{The rotation supported disk can also be thought of as freely levitating.}, but is unlike the pressure or density profiles of any ordinary stellar or planetary atmospheres, which at their base decline linearly with the height.

In the thin atmosphere approximation neither the temperature nor the second derivative of the effective potential can be extracted individually from the pressure or density profiles, they only enter through their ratio $\Phi''(r_0)/T_0$. Note that close to the zero-gravity sphere the pressure has a quasi-universal radial profile depending only on the central temperature to lowest order in height (equation~(\ref{pe})). The polytropic index can only be recovered---once $\Phi''(r_0)/T_0$ is known---near the base of the atmosphere from the density profile (equation~(\ref{rho})), or in the ``upper'' atmosphere (i.e. close to its edges at $r_\pm$) from the full pressure profile  (equation~(\ref{pthing}), or (\ref{pthin})).


\subsection{Exact solutions for the atmospheres}

Maintaining the test-fluid assumption we now focus on the full solutions, in which the atmospheres are not necessarily geometrically thin and their properties then depend on the spacetime under consideration.

We may solve equation~(\ref{eq:hydrostaicEquilibrium}) exactly, once we have a barotropic equation of state $p(\rho)$. The atmospheric profile will depend, in general, on the equation of state, on the metric function $\Phi$, and on the maximum value of pressure $p_0$.

Since we can neglect the pressure term when compared with the energy density, the total mass of the atmosphere is given by 
\begin{equation}
M_{\rm atm} = 4\pi \int \rho\, r^2 e^{\Lambda/c^2} dr,
\label{ADM}
\end{equation}
with the integral evaluated over its whole radial range. In order to be consistent with the test-fluid approximation, one must have $M_{\rm atm}\ll M_{\rm ADM}$, the Arnowitt-Deser-Misner (ADM) mass of the background spacetime.

\begin{center}
\begin{figure*}
\includegraphics[width = 0.95\columnwidth]{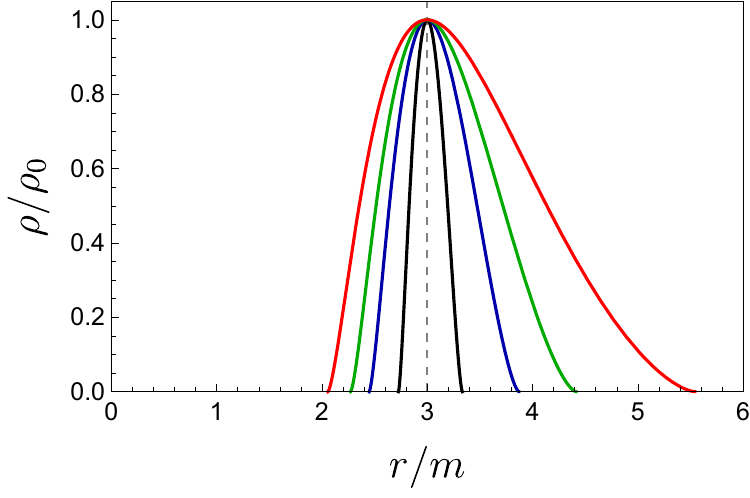}\qquad
\includegraphics[width = 0.95\columnwidth]{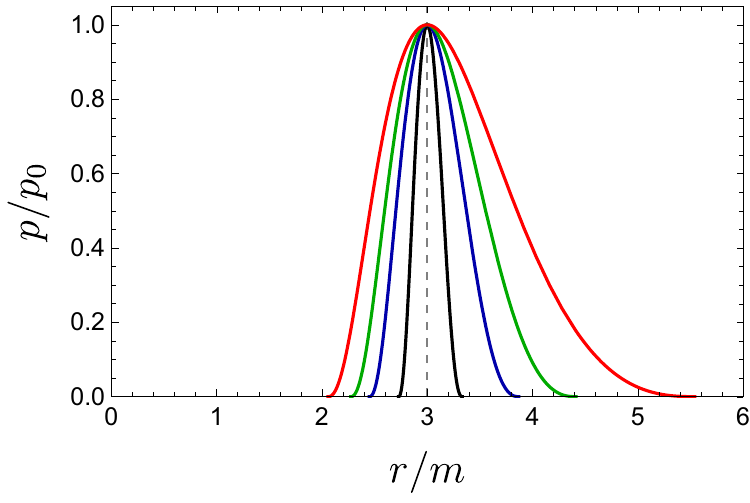}
\caption{Density (left) and pressure (right) profiles for a polytropic atmosphere in Reissner-Nordstr\"om spacetime with $r_0/m=(Q/m)^2 = 3$ and $\gamma = 5/3$.
The curves are given in terms of the parameter $\beta$ defined by   $k_{\rm B}\,T_0 = \beta\,\mu\,m_{\rm p}\,c^2$, with the central temperature given by equation~(\ref{T0}).    From bottom to top: black curves: $\beta=0.001$, blue curves: $\beta=0.005$, green curves: $\beta=0.01$, red curves: $\beta=0.02$. We see that for sub-MeV temperatures (corresponding to $\beta < 0.001$) the atmosphere is fairly symmetric -- see the black curve -- whereas for higher temperatures, of already a few MeV, the density and pressure distribution are skewed owing to the asymmetry of $\Phi(r)$ with respect to the zero-gravity radius.}
\label{fig:rhopRN}
\end{figure*}
\end{center}

\subsubsection{Isothermal atmosphere}

For an ideal gas at constant temperature $T$, we have the profile
\begin{equation}
p(r) = p_0\exp\left\{-\frac{\mu\, m_\mathrm{p}}{\,k_\mathrm{B} T\,}\,\Big[\Phi(r)-\Phi(r_0)\Big]\right\},
\end{equation}
where $p_0>0$ is the maximum pressure; $\rho(r)$ is then given by the ideal gas law. The isothermal atmospheres have infinite extent and $p(r\to\infty)>0$, and then $\rho(r\to\infty)>0$, so they also have an infinite mass. Therefore the solution is unphysical for large radii.
However, the above expression can only be applied to regions of the atmosphere whose temperature can be treated as constant, constraining its range of validity and thus justifying its relevance in the present context.


\subsubsection{Polytropic atmospheres}

If we consider instead an ideal gas satisfying a polytropic EOS of the form  $p = K\rho^\gamma$, we find the solution
\begin{equation}\label{eq:Ppolytropic}
p(r) = p_0\left\{1 - \left(\frac{1-1/\gamma}{K^{1/\gamma}\,p_0^{(1-1/\gamma)}}\right)\,\Big[\Phi(r)-\Phi(r_0)\Big]\right\}^{\gamma/(\gamma-1)},
\end{equation}
where $p_0 = p(r_0)$ is the maximum pressure.
As $p_0$ increases for a fixed value of $K$, the atmosphere gets thicker because $\Phi$ increases with distance from $r_0$. For a given $K$ there is an upper bound, $p_{0,\rm max}$, of the pressure maxima for which the atmosphere does not extend to infinity. 
Since $\Phi(r)>\Phi(r_0)$ in general, then for $p_0 < p_{0,\rm max}$ we may find two points at which $p=0$, one with $r_-<r_0$ and the other with $r_+>r_0$. Then we will have a levitating atmosphere around the naked singularity in the region $r_-<r<r_+$ and delimited by sharp edges where $p=0$ and $\rho=0$.
Each naked singularity spacetime will have a different function $\Phi$. Therefore the atmospheric profile will depend on the spacetime, on the polytropic index $\gamma$ and, apparently, on the parameters $K$ and $p_0$. In fact, we will show that once the spacetime and the polytropic index are fixed, it is the maximum temperature $T_0$ that determines the maximum and minimum radii of the atmosphere.

Indeed, there is also a limiting value $(p_{0}/\rho_{0})_{\rm max}$ allowed for the peak ratio of pressure to density in spatially finite atmospheres, and hence a limiting central temperature by equation~(\ref{cenT}), implicitly given by this expression for the upper bound under discussion:
\begin{equation}
p_{0,\rm max}(K) = \left\{\bigg(\frac{1-1/\gamma}{K^{1/\gamma}}\bigg)\,\Big[-\Phi(r_0)\Big]\right\}^{\gamma/(\gamma-1)},
\label{p0max}
\end{equation}
which yields a condition that does not depend on the polytropic constant $K$,
\begin{equation}
(p_{0}/\rho_{0})_{\rm max}= \bigg(\frac{\gamma-1}{\gamma}\bigg)\,\Big[-\Phi(r_0)\Big].
\label{pmax}
\end{equation}

The density profile $\rho(r)$ is obtained via the polytropic EOS and is given by
\begin{equation}\label{eq:Rhopolytropic}
\rho(r)=\rho_0\left\{1 - \left(\frac{1-1/\gamma}{K\,\rho_0^{(\gamma-1)}}\right)\,\Big[\Phi(r)-\Phi(r_0)\Big]\right\}^{1/(\gamma-1)},
\end{equation}
with $\rho_0 = (p_0/K)^{1/\gamma}$, or
\begin{equation}
\rho(r) = \rho_0\left\{ 1 - \frac{\gamma -1}{\gamma}\, \left(\frac{\mu\, m_\mathrm{p}\,c^2}{k_\mathrm{B}T_0}\right) \left[\frac{\,\Phi(r)}{c^2}-\frac{\Phi(r_0)}{c^2}\right]\right\}^{1/(\gamma-1)}.
\label{rhoT0}
\end{equation}
Similarly, pressure can be expressed as
\begin{equation}
p(r) = p_0\left\{ 1 - \frac{\gamma -1}{\gamma}\, \left(\frac{\mu\, m_\mathrm{p}\,c^2}{k_\mathrm{B}T_0}\right) \left[\frac{\,\Phi(r)}{c^2}-\frac{\Phi(r_0)}{c^2}\right]\right\}^{\gamma/(\gamma-1)}.
\label{pT0}
\end{equation}
In order to have a spatially finite atmosphere, for a fixed value $K$ we must have $\rho_0< \rho_{0,\rm max}(K)$, where
\begin{equation}
\rho_{0,\rm max} (K)= 
\left\{\left(\frac{\gamma-1}{\gamma K}\right)\,\Big[-\Phi(r_0)\Big]\right\}^{1/(\gamma-1)},
\end{equation}
which again translates into condition~(\ref{pmax}). Of course,  for $\gamma>1$, $p_{0,\rm max}(K)/\rho_{0,\rm max}(K)=(p_{0}/\rho_{0})_{\rm max}$ of equation~(\ref{pmax}).

The temperature profile is given by the ideal gas equation and, for the polytropic atmosphere, it reads 
\begin{equation}\label{eq:Tpolytropic}
T(r) = T_0\left\{ 1 - \frac{\gamma -1}{\gamma}\, \left(\frac{\mu\, m_\mathrm{p}\,c^2}{k_\mathrm{B}T_0}\right) \left[\frac{\,\Phi(r)}{c^2}-\frac{\Phi(r_0)}{c^2}\right]\right\}
\end{equation}
 where $p_0/\rho_0 = k_\mathrm{B} T_0/(\mu\, m_\mathrm{p})$. 

The upper bound to allowed central temperatures in a finite polytropic atmosphere, again corresponding to  condition~(\ref{pmax}), is
\begin{equation}
\label{eq:Tmax}
 k_\mathrm{B}T_{0,\rm max}=(\mu\, m_\mathrm{p}c^2)\left(\frac{\gamma-1}{\gamma}\right)\,\Big[-\frac{\Phi(r_0)}{c^2}\Big].
\end{equation}
For the Reissner-Nordstrom metric case illustrated in Fig.~\ref{fig:PhiRN} this is about 0.1 GeV, or $T_0\approx 10^{12}$K, corresponding to $p\approx 0.1\rho c^2$, i.e., it is close to the limit of our assumed range of pressures $p\ll\rho c^2$.
This value of $T_{0,\rm max}$ is much larger than the central temperature of ordinary stars, undergoing nuclear fusion in their interiors. Hence, there seems to be no reason why sufficiently massive hydrogen shells around a naked singularity should not be able to support nuclear fusion, and thus have the external appearance of an ordinary star.


\section{Naked-singularity spacetimes}
\label{sec:nakedSpacetimes}
To plot the radial profiles and discuss the surface density of the atmosphere given in Section~\ref{sec:atmospheres} we need a specific potential function $\Phi$, so we need to turn to a specific metric.

\subsection{Reissner-Nordstr\"om naked singularity}

We will use the Reissner-Nordstr\"om (RN) spacetime as a generic example of naked-singularity spacetimes. In this section we use geometrized units ($G=1=c$). The RN metric is given by

\begin{equation}
\Phi
=\frac{1}{2}\log\left[ 1 - \frac{2m}{r} + \frac{Q^2}{r^2}\right]
\end{equation}
with $\Lambda = - \Phi$. 
Here $m$ is the mass and $Q$ is the charge of the central object. If $Q/m<1$ we have a black hole; $Q/m=1$ gives us an extremal black hole. On the other hand, if $Q/m>1$, the central object is a naked singularity. In this regime, a zero-gravity sphere exists \citep{puglieseQuevedoRuffini2011PRD, katkaVieiraEtal2015GRG}, its radius $r_0$ being equal to $r_0 = Q^2/m$. The second derivative at this {critical point} of the potential is given by
\begin{equation}
r_0^2\Phi''(r_0)=\frac{m}{r_0}\left(1-\frac{m}{r_0}\right)^{-1}.
\label{eq:phibis}
\end{equation}
{Since $r_0/m>1$ in the naked-singularity regime, we indeed have $\Phi''(r_0)>0$ always, leading to stable equilibrium.}

It is interesting to note that as $q\equiv Q/m$ increases the spacetime successively loses its black hole vestiges (Fig.~\ref{fig:stabRN}). First, the horizon disappears (as soon as $q>1$), then the photon orbit, and finally the marginally stable orbit (for $q^2>1.25$, \citealp{puglieseQuevedoRuffini2011PRD}) so that stable circular orbits extend all the way down to the zero-gravity sphere at $r_0$. 
As a specific numerical example of a naked singularity in the latter regime we will take $(Q/m)^2=3$, hence $r_0=3m$, and \begin{equation}
r_0^2\Phi''(r_0)=1/2.
\label{ibis}
\end{equation}

With this value of $r_0$ we can give specific illustrations of the formulae given in the previous Section for the radial profiles of levitating atmospheres, whether exact solutions or in the geometrically thin approximation. Since purely isothermal atmospheres are unphysical, we consider below only the polytropic case.

\subsubsection{Polytropic RN atmospheres}

The polytropic atmospheres around a RN singularity with $r_0/m = (Q/m)^2 = 3$ for a polytropic index $\gamma = 5/3$ are plotted in Fig.~\ref{fig:rhopRN} as functions of the dimensionless parameter $\beta$, defined by $k_{\rm B}\,T_0 = \beta\,\mu\,m_{\rm p}\,c^2$.  E.g., for $\beta=10^{-3}$ (and $\mu = 0.5$) we have $k_{\rm B}\,T_0 \approx 1\,$MeV. Since equations~(\ref{rhoT0}) and (\ref{pT0}) scale with $\rho_0$ and $p_0$, respectively, the plotted dimensionless density and pressure are functions of $\beta$ alone. 
Evidently, the width of the atmospheres increases with temperature (and consequently with $\beta$).

We see from both panels of Fig.~\ref{fig:rhopRN} that for sub-MeV temperatures (corresponding to $\beta < 0.001$) the atmosphere is fairly symmetric -- see the black curve -- whereas for higher temperatures, of already a few MeV, the density and pressure distribution are skewed due to the asymmetry of $\Phi(r)$ with respect to its minimum, the zero-gravity radius.\footnote{Usually, $\Phi(r)$ grows faster in the direction of the centre than in the outer direction (as in Fig.~\ref{fig:PhiRN} for Reissner-Nordstr\"om naked singularity).}

\begin{center}
	\begin{figure*}
		\includegraphics[width = 0.95\columnwidth]{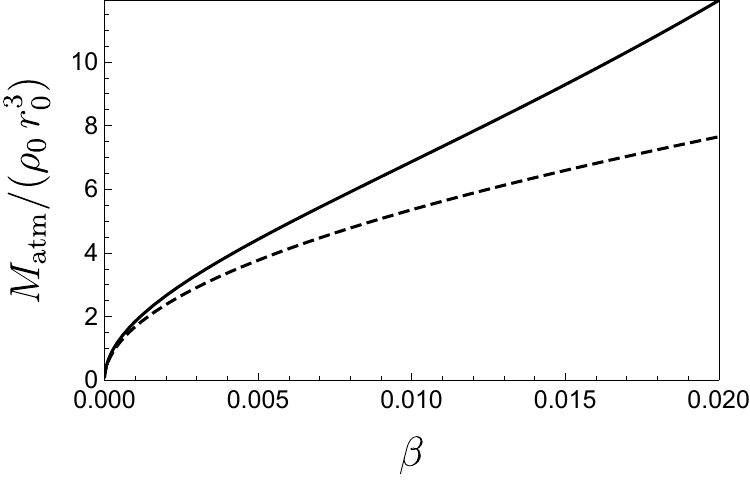}\qquad
		\includegraphics[width = 0.95\columnwidth]{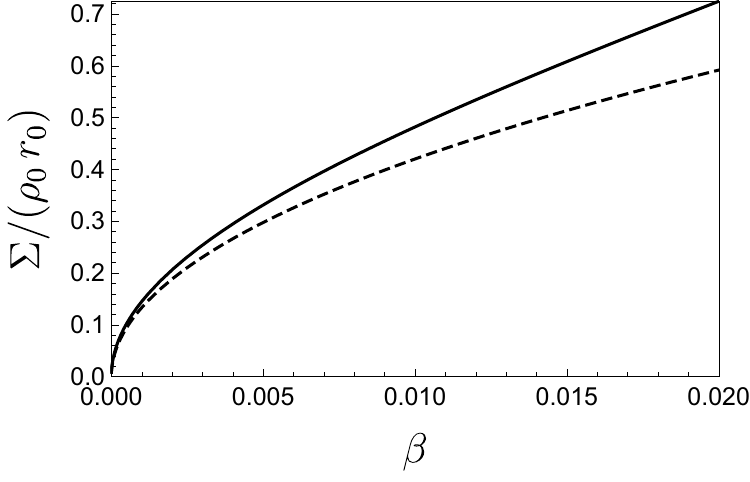}\\ \hfill\\
		\includegraphics[width = 0.95\columnwidth]{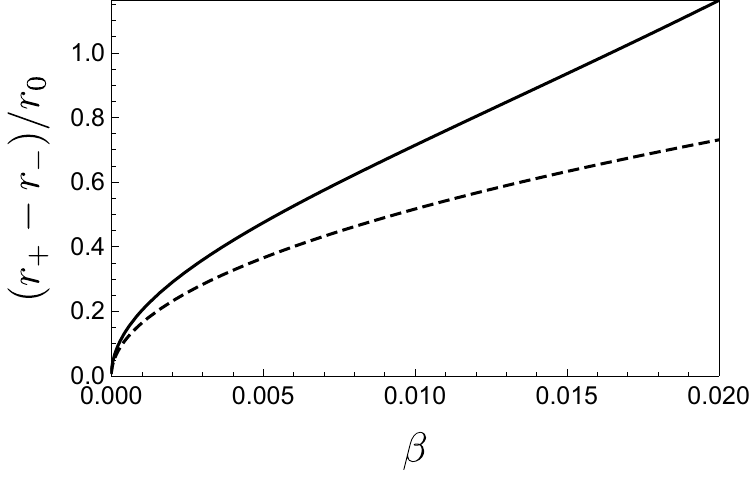}\qquad
        \includegraphics[width = 0.95\columnwidth]{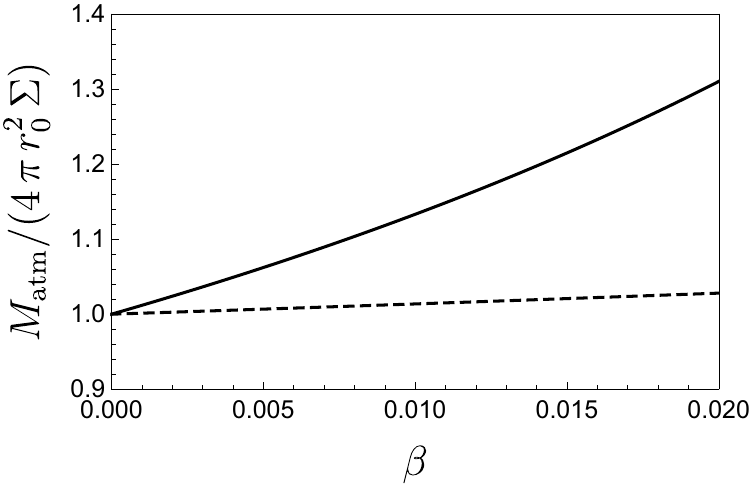}
		\caption{{\sl Top left:} total mass for atmospheres in Reissner-Nordstr\"om spacetime with the parameters of Fig.~\ref{fig:rhopRN} (so $r_0=3m G/c^2$), as a function of the dimensionless central temperature, $\beta=k_{\rm B}\,T_0/(\mu\,m_{\rm p}\,c^2)$. The continuous line corresponds to the exact solution, whereas the dashed line corresponds to the geometrically thin approximation. The maximum plotted value $\beta=0.02$ corresponds to the largest value of the curves in Fig.~\ref{fig:rhopRN}. 
        {\sl Top right:} column density (equation~(\ref{column})) of the corresponding atmospheres as a function of $\beta$. 
        {\sl Bottom left:} geometric thickness of the atmospheres as a function of $\beta$. We note that the behaviour of the thickness, and $\Sigma$, for $\beta \ll 1$ is consistent with equation~(\ref{rminmax}), the ordinate of the curves being proportional to $\beta^{1/2}$. {\sl Bottom right:} The ratio of the actual atmospheric mass to  $4\pi r_0^2 \Sigma$, its thin-surface-layer approximation of equation~(\ref{column0}).
  }
		\label{fig:sigmaRN}
	\end{figure*}
\end{center}

In Fig.~\ref{fig:sigmaRN} we plot, for the same example (RN singularity with $r_0=3m$) the total radial extent of the polytropic atmosphere, $r_+-r_-$, in units of $r_0$, and the column density (density ``height''  
integrated over the radial extent)
\begin{equation}
\Sigma =  \int^{r_+}_{r_-} \rho\,  e^{\Lambda/c^2} dr,
\label{column}
\end{equation}
as well as the total mass (volume integrated) of the atmosphere, equation~(\ref{ADM}). 
From an inspection of the Figures it is clear that the geometrically thin approximation underestimates all these quantities, but only very slightly for temperatures of a few MeV, and in fact imperceptibly for $\beta\ll 0.001$, the sub-MeV temperature regime. 

In the range of $\beta$ plotted---i.e. for temperatures roughly $<20\,$MeV---the test-fluid approximation, $M_{\rm atm}\ll m$, is satisfied as long as the maximum atmospheric density $\rho_0$ is less than the ``mean density'' of the singularity within its zero-gravity surface, $\bar\rho \equiv 3m/(4\pi r_0^3)\,\propto 1/m^2$. Since for a 2 Solar mass singularity $\bar\rho= 1\cdot 10^{15}\,$g/cm$^3$, and for a $4\cdot10^6$ Solar mass singularity $\bar\rho=3\cdot10^2\,$g/cm$^3$, the atmosphere is safely in the test-fluid approximation as long as its density does not exceed nuclear density in the former case (typical low-mass X-ray binary) and the density of water in the latter (Sgr A*).

Note that by equation~(\ref{ibis}) and equation~(\ref{rminmax}) with sub-MeV central temperature, ${k_\mathrm{B}T_0}/({\mu\, m_\mathrm{p}\,c^2})< 10^{-3}$,
the atmosphere is thin: \mbox{$r_+ - r_0 < 0.1\, r_0$}, so that within the atmosphere $\Phi(r)$  differs by at most a few percent from $\Phi(r_0)$ as the lowest order correction is $\Phi(r_0)'' (r-r_0)^2/2 = (r-r_0)^2/(4r_0^2)$. By the same token,
\begin{equation}
 M_{\rm atm} \approx 4\mathrm{\pi}r_0^2\Sigma
\label{column0}
\end{equation}
to within a few percent at most for $\beta\ll 10^{-2}$ (c.f. Fig.~\ref{fig:sigmaRN}, bottom right panel), as expected for a very thin surface layer spread on a sphere of radius $r_0$.

We also verified that the smaller $r_0$ is, the better the geometrically thin approximation. Therefore the illustrative choice $r_0=3m$ is a good example for a compact atmosphere.

\subsection{Other naked singularities}
\label{others}
{It is remarkable that there are space-times in alternative theories of gravity in which the functional form of the metric is exactly the same as that of the RN metric. For instance, Horndeski's scalar-tensor gravity \citep{horndeski1974}, and Randal-Sundrum brane-world model \citep{randallSundrum1999} have this property. Other spacetimes, while having a different metric, preserve the same qualitative features as the RN space-time, illustrated in Fig.~\ref{fig:stabRN}. Thus, as long as we do not discuss the specific effects of electric charge, our discussion should apply to several non-RN singularities.}

For instance, the Kehagias-Sfetsos (KS) spacetime is a spherically symmetric black-hole solution of a modified version of Ho\v{r}ava gravity, which became very popular in the end of the last decade \citep{kehagiasSfetsos2009PhLB}. It is given by 
\begin{equation}
\Phi =\frac{c^2}{2}\log\left\{ 1 + r^2\Omega\left[1-\sqrt{1+\frac{4\,m}{\Omega \,r^3}}\,\right]\, \right\},
\end{equation}
with $\Lambda = - \Phi$. It has an additional parameter $\Omega$ which is intrinsic to the theory. 
If $\Omega m^2 > 1/2$, we have a black-hole solution. For $\Omega m^2 = 1/2$ we have an extremal black hole (with zero surface gravity), whereas for $\Omega m^2 < 1/2$ the central object is a naked singularity.

 As in the RN solution, whenever there is a naked sigularity (i.e., in the parameter range $\Omega m^2 < 1/2$),  a zero-gravity radius exists 
at $r_{0\mathrm{G}}=m(2\Omega m^2)^{-1/3}$, and the stability diagram for timelike circular geodesics  has the same topology as Fig.~\ref{fig:stabRN} \citep{vieiraMarekEtal2014PRD}.
{We refer the reader to that paper for a detailed discussion of the unusual properties of test-particle orbits in naked singularity spacetimes.}
Matter in this theory satisfies the usual conservation laws \citep{germaniKehagiasSfetsos2009JHEP}, so we can construct levitating atmospheres around this naked singularity via the approach presented above. 

While not all space-time metrics have been investigated in equal detail regarding the possible orbital motions, it seems that the existence of a zero-gravity radius is fairly generic {(R. Mishra, in preparation)}.

\section{Atmospheric opacity}
\label{sec:opacity}

The atmosphere will be opaque when it exceeds a certain minimum mass, $M_{\rm atm} > M_{\rm opaque}$. Specifically, the electron scattering optical depth will exceed unity whenever the column density  exceeds the inverse of the scattering opacity, i.e., when $\Sigma>1\mathrm{g/cm}^2$, translating to a laughably small value of the critical minimum mass of an opaque atmosphere,
\begin{equation}
M_{\rm opaque} = 2\cdot10^{14}\mathrm{g}\left(\frac{m}{10M_\odot}\right)^2 \left(\frac{Q}{\sqrt{3}\,m}\right)^4.
\label{opaque}
\end{equation}
Here, we assumed the approximation of equation~(\ref{column0}) and the  RN metric. Also, $Q/m$ is given in geometrized units,  otherwise the masses are in conventional units. Note that
\begin{equation}
\frac{M_{\rm opaque}}{m} = 10^{-22}\left(\frac{m}{M_\odot}\right)\left(\frac{Q}{m}\right)^4,
\end{equation}
where $M_\odot$ is the Solar mass, so the condition $M_{\rm opaque}/m\ll 1 $  is safely satisfied even for the most massive compact objects, as long as\footnote{A recent limit based on EHT observations \citep{2022ApJ...930L..12E} for Sgr A* in the RN metric is $Q/m<2.32$ \citep{MishraVieira2023}.} $Q<300\,m$, given that the largest observed compact object (astrophysical ``black hole") mass is $<10^{11}{M_\odot}$. However, on the low side of the possible mass range, i.e., for low-mass primordial naked singularities, the question of atmospheric opaqueness is well posed only for $r_0$ large enough that the system is in the classical fluid regime, i.e., for $r_0\gg10^{-8}\,$cm, and for the number $N$ of electrons in the atmosphere suitably large, e.g. $N\equiv x M_{\rm atm}/m_\mathrm{p}\gtrsim 10^3$,  the first constraint being more stringent. Hence, our discussion is applicable to $m\gg2\cdot10^{20}\mathrm{g}=10^{-13}M_\odot$. Here, $x\sim1$ is the mean number of electrons per proton, depending on the composition as $Z/A$.

It is hard to imagine an astrophysical context in which a naked singularity would {\sl not} have acquired $M_{\rm opaque}$ through accretion. For example the compact objects in typical X-ray binaries accrete at the rate of 
$10^{-9}M_\odot/$yr in the case of putative $10M_\odot$ black holes, and an order of magnitude higher for $\sim 2M_\odot$ neutron stars. The amount of $M_{\rm opaque}$ is then accreted in about a millisecond or less. In AGNs (active galactic nuclei) the accretion rate is on the order of $10^{20}$g/s, or more, while the compact object mass has $m\sim10^9M_\odot$. Here, the amount $M_{\rm opaque}$ is accreted in less than a thousand years. 

Finally, consider a primordial singularity moving through cold interstellar matter of density $10^{-24}\mathrm{g/cm}^3$ (corresponding to $n=1\,$atom/cm$^3$) at the characteristic speed of $\varv=30\,$km/s, hence sweeping out mass at the rate of $\rho \varv=3\times 10^{-18}\mathrm{g/(cm^2s)}.$ The effective gravitational cross-section of the object is $\mathrm{\pi}R^2$, with $R\sim Gm/\varv^2$, so the mass accretion rate will be 
\begin{equation}
    \dot M=\mathrm{\pi}R^2\rho \varv= 2\cdot10^9\mathrm{({g}/{s})}\left(\frac{m}{M_\odot}\right)^2\left(\frac{\varv}{30\,\mathrm{km/s}}\right)^{-3}\left(\frac{\rho}{m_\mathrm p/\mathrm{cm^3}}\right),
\end{equation}
yielding $M_{\rm opaque}$ in about $10^3$ seconds for $\varv=30\,$km/s and $\rho=1 m_\mathrm p/\mathrm{cm^3}$. Note that this time, $\tau_{\rm o} \equiv M_{\rm opaque}/\dot M$, does not depend on the mass of the singularity, as both quantities in the ratio have the same square dependence on $m$ (for a fixed $Q/m$ value). 

Thus, in all three scenarios the time to cloak the singularity by accreted matter is extremely short by astronomical standards.
\section{Atmospheric oscillations}
\label{sec:Oscillation}
Near the minimum, the potential $\Phi$ can be approximated by a parabola
$\Phi(r)= \Phi(r_0) + \Phi''(r_0) (r-r_0)^2/2$
and therefore any body placed close to $r_0$ will execute harmonic oscillations at the proper angular frequency $\omega_{\mathrm{p}} = \exp[{-\Lambda(r_0)/c^2}]\,\sqrt{\Phi''(r_0)}.$ The frequency observed at infinity is $\omega=\exp[-{\Phi(r_0)/c^2}]\omega_{\mathrm{p}} $, and since $\Lambda=-\Phi$ in the RN metric the observed frequency will be
\begin{equation}
\omega= \sqrt{\Phi''(r_0)}.
\label{eq:qpo}
\end{equation} 
This is also true for the thin atmosphere as a whole. Small amplitude harmonic motion of the thin atmosphere at the  frequency $\omega$ is an exact eigenmode of this fluid body. In this fundamental mode the atmosphere as a whole executes uniform, spherically symmetric harmonic radial displacements from the equilibrium position. The first overtone corresponds to the breathing mode, at frequency $\sqrt{\gamma+1}\,\omega$. For a discussion of this point see \citet{bollimpalliKluzniak2017MNRAS}, especially their equation~(18).

Being supported by conservative gravity, the oscillating atmosphere will only be weakly damped by dissipation of internal motions. The fundamental mode of oscillation, being a rigid-body motion in the geometrically thin atmosphere limit, will have very little damping.\footnote{Since we treat the Reissner-Nordstr\"om metric as a generic example of a spherically symmetric naked singularity, we only discuss gravitational effects and ignore the direct electromagnetic influence of the electric charge $Q$ (such as the electric polarization of the levitating shell).}
 Thus we expect this mode to have a particularly large quality factor. This is unlike the strongly damped ECS oscillations considered by
\citet{abarcaKluzniak2016MNRAS} and \citet{bollimpalliEtal2019MNRAS}, where the atmosphere was supported by highly dissipative forces of external radiation,
rather than by gravity---as is the case here. A long lived oscillation of the atmospheric shell could then perhaps be a signature of a cloaked naked singularity.

For the Reissner-Nordstr\"om singularity, restoring the $G$ and $c^2$ factors in equation~(\ref{eq:phibis}) we obtain
\begin{equation}
\omega^2=\frac{Gm}{r_0^3}\left(1-\frac{Gm}{c^2 r_0 }\right)^{-1}.
\label{eq:qporn}
\end{equation}
For the specific example of $r_0=3Gm/c^2$, we have equation (\ref{ibis}), and the fundamental frequency is about 750 Hz for a $10\,M_\odot$ singularity,
\begin{equation}
f=\frac{\omega}{2\mathrm{\pi}}=\frac{c}{r_0}\left(2\sqrt{2}\,\mathrm{\pi}\right)^{-1}=7.5\, \mathrm{kHz}\left(\frac{m}{M_\odot}\right)^{-1}.
\label{eq:qporn2}
\end{equation}
This frequency is only slightly too high for the oscillations to be associated with the observed high frequency QPOs in X-ray binaries
\citep{2000ARA&A..38..717V}.
A small change in the parameters of the considered naked singularity would bring the predicted frequencies into alignment with those observed in X-ray binaries.

\section{Discussion}
\label{sec:Discussion}

We obtained here analytical solutions for levitating atmospheres around 
spherically symmetric naked singularities. 
There is a stable equilibrium position for radial particle motion due to gravity only, at any point of a spherical surface of a certain radius $r_0$, the ``zero-gravity radius''. 
Matter accreting onto the singularity will eventually be deposited, and come to rest, on this ``zero-gravity sphere'' through the action of dissipative mechanisms in the fluid, thus generating an
atmosphere with density peaking at this radius.
Moreover, since the dynamics is governed by gravity only, the same formalism applies to both optically thin and optically thick atmospheres, the difference being due only to the choice of parameters of the final solution.
We remark that the presence of a singularity is not essential to these solutions. Atmospheres can be constructed (via the same formalism) around any spherically symmetric object which possesses a zero-gravity radius, 
even in the absence of a singularity. Such is the case 
in the no-horizon parameter region of regular black-hole solutions in nonlinear electrodynamics \citep{garciaHackmannEta2015JMP} and in certain theories of modified gravity (see \S~\ref{others}).

We have shown that in typical astrophysical scenarios, in a very short time the naked singularity will accrete enough matter to be opaque to electron scattering, whether the singularity is in a semidetached binary, at the center of a Galaxy, or traversing interstellar matter. Thus, while the singularity will be naked in the sense of not possessing an event horizon, it will in fact be ``cloaked'' by its atmosphere, i.e. its inner (antigravity) regions will be inaccessible to direct observation in the optical or X-ray domain. The external appearance of the cloaked singularity will be similar to that of a planet or star.

We have constructed solutions for isothermal and polytropic atmospheres. Although purely isothermal atmospheres are unphysical since they extend to infinity, having infinite mass \citep[a fact that also happens in luminous neutron stars,][]{wielgusEtal2015MNRAS}, the expression obtained here can be used as a description of that part of the atmosphere which  has an isothermal behaviour (if it does).
On the other hand, polytropic equations of state give us realistic levitating atmospheres, either optically thin or thick.  The geometrical thickness of the polytropic atmosphere depends on the central temperature of the atmosphere. The optically thick case is particularly interesting, since the atmosphere would be responsible for ``cloaking'' the naked singularity. 
Moreover, the polytropic atmospheres may be very compact.

As the  central temperature of the atmospheres increases, their radial extent increases; there is a maximum peak temperature which guarantees finite atmospheres. For larger temperatures, the atmospheres have infinite extent and infinite mass, becoming unphysical.
While this gives an upper limit on the central temperature of the levitating atmospheres, it is so large ($\sim10^{12}$K) as not to be restrictive. We see no reason why a levitating atmosphere may not become sufficiently dense and hot to ignite hydrogen at its base. Thus, to the outside world, a levitating atmosphere may take on the appearance of a common star, even though in fact it would be a shell, hollow inside.

Through measurements of spectral line broadening, astronomers can determine the surface gravity $g_\mathrm{surf}$ of a star \citep[e.g.,][]{2014dapb.book...97C}, which of course gives a constraint on the mass and radius through $g_\mathrm{surf}\approx GM_*/R_*^2$. The radius of the photosphere, $R_*$ may directly be measured from the luminosity and effective temperature of the star. In a binary, the mass $M_*$ may be measured independently. If similar measurements were possible for a fairly thin levitating atmosphere around a naked singularity, a surprising result may be obtained that $g_\mathrm{surf}\ll GM_*/R_*^2$. Indeed, at $r_0$ the effective gravity is zero, and it increases gradually with radius, so that at $R_*\approx r_0$ there is no simple (universal) relationship of $g_\mathrm{surf}$ to the mass $m=M_*$ of the singularity, as measured ``at infinity'' through the motion of the binary components, for instance. 

Since the atmospheres are supported only by gravity,
their radial modes will not be strongly damped; therefore, the long-term oscillations of such atmospheres, should they be excited, could be a signature of a cloaked singularity.
We postpone a detailed discussion of radial modes to a future investigation.

Finally, we speculate  that naked singularities may provide an excellent scaffolding for creating low-mass neutron stars, or strictly speaking their {``hollow''} analogues. As is clear from  the discussion of Fig.~\ref{fig:sigmaRN} and equation~(\ref{column}), for $m=M_\odot$ a maximum atmospheric density as high as $4\cdot10^{15}\,\mathrm{g/cm^3}$ satisfies the test fluid constraint, so clearly, supranuclear atmospheric densities are within the scope of our discussion for $m\lesssim1M_\odot$ naked singularities. In such a singularity  an atmosphere of only a fraction of a Solar mass may already be compressed to supranuclear densities and may form neutron-star matter. Assuming that spinning up such an object would primarily involve the low-mass shell, and depending on the radius $r_0$ of the zero-gravity sphere, the moment of inertia of such a star could be much lower in proportion to its mass than that of a true neutron star of the same total mass that is not hollow inside.

%
\section*{Acknowledgments}
This work was partially supported by the Coordena\c{c}\~ao de Aperfei\c{c}oamento de Pessoal de N\'ivel Superior - Brasil (CAPES) - Finance Code 001, under the Brazilian CAPES-PrInt internationalization program, and supported in part by the
Polish NCN grant 2019/33/B/ST9/01564.
%

%
\section*{Data Availability}
No new data were generated or analysed in support of this research.





\begin{thebibliography}{}
	\makeatletter
	\relax
	\def\mn@urlcharsother{\let\do\@makeother \do\$\do\&\do\#\do\^\do\_\do\%\do\~}
	\def\mn@doi{\begingroup\mn@urlcharsother \@ifnextchar [ {\mn@doi@}
		{\mn@doi@[]}}
	\def\mn@doi@[#1]#2{\def\@tempa{#1}\ifx\@tempa\@empty \href
		{http://dx.doi.org/#2} {doi:#2}\else \href {http://dx.doi.org/#2} {#1}\fi
		\endgroup}
	\def\mn@eprint#1#2{\mn@eprint@#1:#2::\@nil}
	\def\mn@eprint@arXiv#1{\href {http://arxiv.org/abs/#1} {{\tt arXiv:#1}}}
	\def\mn@eprint@dblp#1{\href {http://dblp.uni-trier.de/rec/bibtex/#1.xml}
		{dblp:#1}}
	\def\mn@eprint@#1:#2:#3:#4\@nil{\def\@tempa {#1}\def\@tempb {#2}\def\@tempc
		{#3}\ifx \@tempc \@empty \let \@tempc \@tempb \let \@tempb \@tempa \fi \ifx
		\@tempb \@empty \def\@tempb {arXiv}\fi \@ifundefined
		{mn@eprint@\@tempb}{\@tempb:\@tempc}{\expandafter \expandafter \csname
			mn@eprint@\@tempb\endcsname \expandafter{\@tempc}}}
	
	\bibitem[\protect\citeauthoryear{{Abarca} \& {Klu{\'z}niak}}{{Abarca} \&
		{Klu{\'z}niak}}{2016}]{abarcaKluzniak2016MNRAS}
	{Abarca} D.,  {Klu{\'z}niak} W.,  2016, \mn@doi [\mnras]
	{10.1093/mnras/stw1432}, \href
	{https://ui.adsabs.harvard.edu/abs/2016MNRAS.461.3233A} {461, 3233}
	
	\bibitem[\protect\citeauthoryear{{Abramowicz}, {Ellis}  \&
		{Lanza}}{{Abramowicz} et~al.}{1990}]{abramowiczEtal10990ApJ}
	{Abramowicz} M.~A.,  {Ellis} G.~F.~R.,   {Lanza} A.,  1990, \mn@doi [\apj]
	{10.1086/169211}, 361, 470
	
	\bibitem[\protect\citeauthoryear{{Bollimpalli} \& {Klu{\'z}niak}}{{Bollimpalli}
		\& {Klu{\'z}niak}}{2017}]{bollimpalliKluzniak2017MNRAS}
	{Bollimpalli} D.~A.,  {Klu{\'z}niak} W.,  2017, \mn@doi [\mnras]
	{10.1093/mnras/stx2140}, \href
	{https://ui.adsabs.harvard.edu/abs/2017MNRAS.472.3298B} {472, 3298}
	
	\bibitem[\protect\citeauthoryear{{Bollimpalli}, {Wielgus}, {Abarca}  \&
		{Klu{\'z}niak}}{{Bollimpalli} et~al.}{2019}]{bollimpalliEtal2019MNRAS}
	{Bollimpalli} D.~A.,  {Wielgus} M.,  {Abarca} D.,   {Klu{\'z}niak} W.,  2019,
	\mn@doi [\mnras] {10.1093/mnras/stz1597}, \href
	{https://ui.adsabs.harvard.edu/abs/2019MNRAS.487.5129B} {487, 5129}
	
	\bibitem[\protect\citeauthoryear{{Boshkayev}, {Gasper{\'{\i}}n},
		{Guti{\'e}rrez-Pi{\~n}eres}, {Quevedo}  \& {Toktarbay}}{{Boshkayev}
		et~al.}{2016}]{boshkakayev2016PRD}
	{Boshkayev} K.,  {Gasper{\'{\i}}n} E.,  {Guti{\'e}rrez-Pi{\~n}eres} A.~C.,
	{Quevedo} H.,   {Toktarbay} S.,  2016, \mn@doi [\prd]
	{10.1103/PhysRevD.93.024024}, \href
	{http://adsabs.harvard.edu/abs/2016PhRvD..93b4024B} {93, 024024}
	
	\bibitem[\protect\citeauthoryear{{Catanzaro}}{{Catanzaro}}{2014}]{2014dapb.book...97C}
	{Catanzaro} G.,  2014, in , Determination of Atmospheric Parameters of B-, A-,
	F- and G-Type Stars. Niemczura, E., Smalley, B., Pych, W. (eds).
	pp 97--109, \mn@doi{10.1007/978-3-319-06956-2_9}
	
	\bibitem[\protect\citeauthoryear{{Christodoulou}}{{Christodoulou}}{1984}]{christodoulou1984}
	{Christodoulou} D.,  1984, \mn@doi [Communications in Mathematical Physics]
	{10.1007/BF01223743}, \href
	{https://ui.adsabs.harvard.edu/abs/1984CMaPh..93..171C} {93, 171}
	
	\bibitem[\protect\citeauthoryear{{Eardley} \& {Smarr}}{{Eardley} \&
		{Smarr}}{1979}]{eardleySmarr1979}
	{Eardley} D.~M.,  {Smarr} L.,  1979, \mn@doi [\prd] {10.1103/PhysRevD.19.2239},
	\href {https://ui.adsabs.harvard.edu/abs/1979PhRvD..19.2239E} {19, 2239}
	
	\bibitem[\protect\citeauthoryear{{Event Horizon Telescope Collaboration}
		et~al.,}{{Event Horizon Telescope Collaboration}
		et~al.}{2022}]{2022ApJ...930L..12E}
	{Event Horizon Telescope Collaboration} et~al., 2022, \mn@doi [\apjl]
	{10.3847/2041-8213/ac6674}, \href
	{https://ui.adsabs.harvard.edu/abs/2022ApJ...930L..12E} {930, L12}
	
	\bibitem[\protect\citeauthoryear{{Garc{\'{\i}}a}, {Hackmann}, {Kunz},
		{L{\"a}mmerzahl}  \& {Mac{\'{\i}}as}}{{Garc{\'{\i}}a}
		et~al.}{2015}]{garciaHackmannEta2015JMP}
	{Garc{\'{\i}}a} A.,  {Hackmann} E.,  {Kunz} J.,  {L{\"a}mmerzahl} C.,
	{Mac{\'{\i}}as} A.,  2015, \mn@doi [Journal of Mathematical Physics]
	{10.1063/1.4913882}, \href
	{http://adsabs.harvard.edu/abs/2015JMP....56c2501G} {56, 032501}
	
	\bibitem[\protect\citeauthoryear{{Germani}, {Kehagias}  \& {Sfetsos}}{{Germani}
		et~al.}{2009}]{germaniKehagiasSfetsos2009JHEP}
	{Germani} C.,  {Kehagias} A.,   {Sfetsos} K.,  2009, \mn@doi [Journal of High
	Energy Physics] {10.1088/1126-6708/2009/09/060}, 9, 60
	
	\bibitem[\protect\citeauthoryear{{Giamb{\`o}}, {Giannoni}, {Magli}  \&
		{Piccione}}{{Giamb{\`o}} et~al.}{2004}]{giamboGiannoniMagliPiccione2004}
	{Giamb{\`o}} R.,  {Giannoni} F.,  {Magli} G.,   {Piccione} P.,  2004, \mn@doi
	[General Relativity and Gravitation] {10.1023/B:GERG.0000022388.11306.e1},
	\href {https://ui.adsabs.harvard.edu/abs/2004GReGr..36.1279G} {36, 1279}
	
	\bibitem[\protect\citeauthoryear{{Goluchov{\'a}}, {Kulczycki}, {Vieira},
		{Stuchl{\'{\i}}k}, {Klu{\'z}niak}  \& {Abramowicz}}{{Goluchov{\'a}}
		et~al.}{2015}]{katkaVieiraEtal2015GRG}
	{Goluchov{\'a}} K.,  {Kulczycki} K.,  {Vieira} R.~S.~S.,  {Stuchl{\'{\i}}k} Z.,
	{Klu{\'z}niak} W.,   {Abramowicz} M.,  2015, \mn@doi [General Relativity and
	Gravitation] {10.1007/s10714-015-1976-3}, 47, 132
	
	\bibitem[\protect\citeauthoryear{Griffiths \& Podolsk{\`y}}{Griffiths \&
		Podolsk{\`y}}{2009}]{griffithsPodolsky2009exact}
	Griffiths J.~B.,  Podolsk{\`y} J.,  2009, Exact space-times in Einstein's
	general relativity.
	Cambridge University Press
	
	\bibitem[\protect\citeauthoryear{{Horndeski}}{{Horndeski}}{1974}]{horndeski1974}
	{Horndeski} G.~W.,  1974, \mn@doi [International Journal of Theoretical
	Physics] {10.1007/BF01807638}, \href
	{https://ui.adsabs.harvard.edu/abs/1974IJTP...10..363H} {10, 363}
	
	\bibitem[\protect\citeauthoryear{{Joshi}}{{Joshi}}{1993}]{joshiBook1993}
	{Joshi} P.~S.,  1993, {Global aspects in gravitation and cosmology}.
	Int. Ser. Monogr. Phys Vol. 87
	
	\bibitem[\protect\citeauthoryear{{Joshi}, {Dadhich}  \& {Maartens}}{{Joshi}
		et~al.}{2002}]{joshiDadhichMaartens2002}
	{Joshi} P.~S.,  {Dadhich} N.,   {Maartens} R.,  2002, \mn@doi [\prd]
	{10.1103/PhysRevD.65.101501}, \href
	{https://ui.adsabs.harvard.edu/abs/2002PhRvD..65j1501J} {65, 101501}
	
	\bibitem[\protect\citeauthoryear{{Joshi}, {Malafarina}  \& {Narayan}}{{Joshi}
		et~al.}{2011}]{joshiMalafarinaNarayan2011}
	{Joshi} P.~S.,  {Malafarina} D.,   {Narayan} R.,  2011, \mn@doi [Classical and
	Quantum Gravity] {10.1088/0264-9381/28/23/235018}, \href
	{https://ui.adsabs.harvard.edu/abs/2011CQGra..28w5018J} {28, 235018}
	
	\bibitem[\protect\citeauthoryear{{Kehagias} \& {Sfetsos}}{{Kehagias} \&
		{Sfetsos}}{2009}]{kehagiasSfetsos2009PhLB}
	{Kehagias} A.,  {Sfetsos} K.,  2009, \mn@doi [Physics Letters B]
	{10.1016/j.physletb.2009.06.019}, 678, 123
	
	\bibitem[\protect\citeauthoryear{{Kov{\'a}cs} \& {Harko}}{{Kov{\'a}cs} \&
		{Harko}}{2010}]{kovacsHarko2010PRD}
	{Kov{\'a}cs} Z.,  {Harko} T.,  2010, \mn@doi [\prd]
	{10.1103/PhysRevD.82.124047}, 82, 124047
	
	\bibitem[\protect\citeauthoryear{{Mishra} \& {Vieira}}{{Mishra} \&
		{Vieira}}{2023}]{MishraVieira2023}
	{Mishra} R.,  {Vieira} R. S.~S.,  2023, 
	\href{https://arxiv.org/abs/2304.04313}{arxiv:2304.04313}
	
	\bibitem[\protect\citeauthoryear{{Oh}, {Kim}  \& {Lee}}{{Oh}
		et~al.}{2010}]{ohEtal2010PhRvD}
	{Oh} J.~S.,  {Kim} H.,   {Lee} H.~M.,  2010, \mn@doi [\prd]
	{10.1103/PhysRevD.81.084005}, 81, 084005
	
	\bibitem[\protect\citeauthoryear{{Ori} \& {Piran}}{{Ori} \&
		{Piran}}{1990}]{oriPiran1990}
	{Ori} A.,  {Piran} T.,  1990, \mn@doi [\prd] {10.1103/PhysRevD.42.1068}, \href
	{https://ui.adsabs.harvard.edu/abs/1990PhRvD..42.1068O} {42, 1068}
	
	\bibitem[\protect\citeauthoryear{{Penrose}}{{Penrose}}{1969}]{penrose1969}
	{Penrose} R.,  1969, Nuovo Cimento Rivista Serie, \href
	{https://ui.adsabs.harvard.edu/abs/1969NCimR...1..252P} {1, 252}
	
	\bibitem[\protect\citeauthoryear{{Pringle}}{{Pringle}}{1981}]{pringle1981ARAA}
	{Pringle} J.~E.,  1981, \mn@doi [\araa] {10.1146/annurev.aa.19.090181.001033},
	19, 137
	
	\bibitem[\protect\citeauthoryear{{Pugliese}, {Quevedo}  \&
		{Ruffini}}{{Pugliese} et~al.}{2011}]{puglieseQuevedoRuffini2011PRD}
	{Pugliese} D.,  {Quevedo} H.,   {Ruffini} R.,  2011, \mn@doi [\prd]
	{10.1103/PhysRevD.83.024021}, 83, 024021
	
	\bibitem[\protect\citeauthoryear{{Randall} \& {Sundrum}}{{Randall} \&
		{Sundrum}}{1999}]{randallSundrum1999}
	{Randall} L.,  {Sundrum} R.,  1999, \mn@doi [\prl]
	{10.1103/PhysRevLett.83.3370}, \href
	{https://ui.adsabs.harvard.edu/abs/1999PhRvL..83.3370R} {83, 3370}
	
	\bibitem[\protect\citeauthoryear{Schutz}{Schutz}{2009}]{schutz2009book}
	Schutz B.,  2009, A first course in general relativity.
	Cambridge university press
	
	\bibitem[\protect\citeauthoryear{{Semer{\'a}k}, {Zellerin}  \& {{\v Z}{\'a}{\v
				c}ek}}{{Semer{\'a}k} et~al.}{1999}]{semerakZellerinZacek1999MNRAS}
	{Semer{\'a}k} O.,  {Zellerin} T.,   {{\v Z}{\'a}{\v c}ek} M.,  1999, \mn@doi
	[\mnras] {10.1046/j.1365-8711.1999.02748.x}, 308, 691
	
	\bibitem[\protect\citeauthoryear{{Singh} \& {Joshi}}{{Singh} \&
		{Joshi}}{1996}]{singhJoshi1996}
	{Singh} T.~P.,  {Joshi} P.~S.,  1996, \mn@doi [Classical and Quantum Gravity]
	{10.1088/0264-9381/13/3/019}, \href
	{https://ui.adsabs.harvard.edu/abs/1996CQGra..13..559S} {13, 559}
	
	\bibitem[\protect\citeauthoryear{{Stahl}, {Wielgus}, {Abramowicz},
		{Klu{\'z}niak}  \& {Yu}}{{Stahl} et~al.}{2012}]{stahlEtal2012AA}
	{Stahl} A.,  {Wielgus} M.,  {Abramowicz} M.,  {Klu{\'z}niak} W.,   {Yu} W.,
	2012, \mn@doi [\aap] {10.1051/0004-6361/201220187}, \href
	{https://ui.adsabs.harvard.edu/abs/2012A&A...546A..54S} {546, A54}
	
	\bibitem[\protect\citeauthoryear{{Stahl}, {Klu{\'z}niak}, {Wielgus}  \&
		{Abramowicz}}{{Stahl} et~al.}{2013}]{stahlEtal2013AA}
	{Stahl} A.,  {Klu{\'z}niak} W.,  {Wielgus} M.,   {Abramowicz} M.,  2013,
	\mn@doi [\aap] {10.1051/0004-6361/201321595}, \href
	{https://ui.adsabs.harvard.edu/abs/2013A&A...555A.114S} {555, A114}
	
	\bibitem[\protect\citeauthoryear{{Vieira}, {Schee}, {Klu{\'z}niak},
		{Stuchl{\'{\i}}k}  \& {Abramowicz}}{{Vieira}
		et~al.}{2014}]{vieiraMarekEtal2014PRD}
	{Vieira} R.~S.~S.,  {Schee} J.,  {Klu{\'z}niak} W.,  {Stuchl{\'{\i}}k} Z.,
	{Abramowicz} M.,  2014, \mn@doi [\prd] {10.1103/PhysRevD.90.024035}, 90,
	024035
	
	\bibitem[\protect\citeauthoryear{{Wielgus}}{{Wielgus}}{2019}]{wielgus2019MNRAS}
	{Wielgus} M.,  2019, \mn@doi [\mnras] {10.1093/mnras/stz2079}, \href
	{https://ui.adsabs.harvard.edu/abs/2019MNRAS.488.4937W} {488, 4937}
	
	\bibitem[\protect\citeauthoryear{{Wielgus}, {Stahl}, {Abramowicz}  \&
		{Klu{\'z}niak}}{{Wielgus} et~al.}{2012}]{wielgusEtal2012AA}
	{Wielgus} M.,  {Stahl} A.,  {Abramowicz} M.,   {Klu{\'z}niak} W.,  2012,
	\mn@doi [\aap] {10.1051/0004-6361/201220228}, 545, A123
	
	\bibitem[\protect\citeauthoryear{{Wielgus}, {Klu{\'z}niak}, {S\c{a}dowski},
		{Narayan}  \& {Abramowicz}}{{Wielgus} et~al.}{2015}]{wielgusEtal2015MNRAS}
	{Wielgus} M.,  {Klu{\'z}niak} W.,  {S\c{a}dowski} A.,  {Narayan} R.,
	{Abramowicz} M.,  2015, \mn@doi [\mnras] {10.1093/mnras/stv2191}, \href
	{https://ui.adsabs.harvard.edu/abs/2015MNRAS.454.3766W} {454, 3766}
	
	\bibitem[\protect\citeauthoryear{{van der Klis}}{{van der
			Klis}}{2000}]{2000ARA&A..38..717V}
	{van der Klis} M.,  2000, \mn@doi [\araa]
	{10.1146/annurev.astro.38.1.71710.48550/arXiv.astro-ph/0001167}, \href
	{https://ui.adsabs.harvard.edu/abs/2000ARA&A..38..717V} {38, 717}
	
	\makeatother
\end{thebibliography}

\bsp	
\label{lastpage}
\end{document}